\documentclass{article}
\usepackage{amsmath}
\usepackage{latexsym}

\title{Strings from Flux Tube Solutions in Kaluza-Klein Theory}
\author{V. Dzhunushaliev
\thanks{E-mail: dzhun@hotmail.kg}}

\date{}

\begin{document}
\maketitle

\begin{center}
\textit{Freie Universit\"at Berlin, Arnimallee 14, D-14195, Berlin, Germany\\
and \\
Dept. Phys. and Microel. Engineer., KRSU \\
Bishkek, Kievskaya Str. 44, 720000, Kyrgyz Republic}
\end{center}

\begin{abstract}
We calculate dimensional reduction of gravitational flux 
tube solutions in the scheme of Kaluza-Klein theory. 
The fifth dimension is compacified to a region of Planck 
size. Assuming the width of the tube to be also Planck size 
we obtain string-like object with physical fields originated from 
an initial 5D metric. The dynamics of these fields is inner one. 
\end{abstract}

\section{Introduction}

In string theory it assumed that strings are line-like objects 
without inner structure. 
Historically, strings arose in an attempt to explain the infinitelly 
rising Regge trajectories of hadrons. In field theory such 
objects were found in non-Abelian gauge theory in the form of 
Nielsen-Olesen flux tubes with a finite thickness. 
We would like to point out that in 5D Kaluza - Klein gravity 
there are vacuum solutions which have the form of strings \cite{dzh2}. 
\par 
Initially, the thickness of a string in quantum field theory 
is arbitrary since the classical theory is scale invariant. 
Quantum fluctuations, however, will fix a scale and string 
obtaines a thickness. 
The finite thickness gives such a string an extrinsic curvature stifness 
\cite{pol} \cite{kleinert} which is absent in the Nambu-Goto string 
Lagrangian. 
\par 
Ordinary, string dynamics arises from a field equations 
for the movement in an external spacetime. Of coarse we do not have 
any another possibility because strings can have only such degrees 
of freedom as their coordinates. One can call such 
string dynamics as an external dynamics. The string-like objects  
derived here have 
an inner structure, and thus an inner dynamics : the dynamics of 
inner degrees of freedom. In this case we do not need for an external 
spacetime for the description of such kind of string dynamics. 
Ordinary, the fields on strings are introduced by hand but if strings 
has an inner structure then some fields will arise by the natural way 
as the consequence of the inner structure. 
\par 
The basic goal of this paper is to show that for string-like objects 
considered here (gravitational flux tube solutions) the fields on 
string have the origin from the 5D metric and the dynamics follows from the 
5D Einstein's equations. 

\section{Gravitational flux tube solutions}

We begin with a brief reminder of the flux tube solutions 
\cite{dzhsin} of the vacuum Kaluza-Klein gravity. The topology of these 
solutions is $M^2 \times S^2 \times S^1$ 
where $M^2$ is the 2D space-time spanned on the time and longitudinal 
coordinate $r$; $S^2$ is the cross section of this flux tube solution and it is 
spanned on the ordinary spherical coordinates $\theta$ and $\varphi$; 
$S^1 = U(1)$ is abelian gauge group which in this consideration is 
the 5$^{\mathrm{th}}$ dimension. 
\par 
The possible reduction from a flux tube to strings 
is based on the following remark : the linear sizes of $S^2$ is not defined from 
Einstein's equations and can be arbitrary. Our strategy is that we set the 
radius of sphere $S^2$ to a Planck length $l_{Pl}$. In QCD the transversal 
sazes of flux tube stretched between quark-antiquark is an inner property : 
in principle it can be calculated inside of QCD. This gives us a possibility 
that after quantum fluctuations are taken into account, the calculated 
thickness of the gravitational flux tube will be of the order of the Planck 
scale. 
\par 
Mathematically the cross section of the flux tube is 
not the point but physically it is indistiquishable with a point because 
one of the paradigms of quantum gravity claimes that the lengths less 
the Planck length have not any sense. Thus, from this physical point of view 
our flux tube solution after this reduction is like to string 
(of coarse we should have the length of the 5$^{\mathrm{th}}$ coordinate 
$\approx l_{Pl}$, too). 
\par
Now we can consider more carefully the flux tube solutions \cite{dzhsin}
in 5D Kaluza-Klein gravity. The 5D metric we take in the form 
\begin{eqnarray} 
ds^2 &=& e^{2\nu (r)}dt^{2} - l_0^2e^{2\psi (r) - 2\nu (r)}
\left [d\chi +  \omega (r)dt + Q \cos \theta d\varphi \right ]^2  
\nonumber\\
&-& dr^{2} - a(r)(d\theta ^{2} + 
\sin ^{2}\theta  d\varphi ^2),
\label{sec1-10}
\end{eqnarray}
where $\chi $ is the 5$^{th}$ extra coordinate; 
$r,\theta ,\varphi$ are $3D$  spherical-polar coordinates; 
$Q$ is some constant; $r \in \{ -R_0 , +R_0 \}$ 
($R_0$ may be equal to $\infty$); $l_0$ is the radius of 5$^{th}$ 
coordinate. $\omega (r)$ is the $t$-component of the electromagnetic 
potential and $Q \cos \theta$ is the $\varphi$-component. 
This means that we have radial Kaluza-Klein 
electrical $E_r$ and magnetic $H_r \propto Q/r^2$ fields. 
The $R_{5\chi}$ equation 
\begin{equation}
\omega '' - 4\nu'\omega' + 3\omega '\psi ' + 
\frac{a'\omega '}{a} = 
\frac{e^{-3\psi + 4\nu}}{4\pi a}
\left(\omega ' e^{3\psi - 4\nu} 4 \pi a \right)' = 0.
\label{sec1-15}
\end{equation}
give us 
\begin{equation}
\label{sec1-16}
E_r = \omega ' = \frac{q}{a} e^{4 \nu - 3 \psi}
\end{equation}
where $q$ is some constant. From the $R_{5\chi}$ equation we see that 
it is like to the Maxwell equation in a continuous medium. In this case 
$e^{3\psi - 4 \nu}$ is like to a dielectric permeability and 
$\mathcal D_r = \omega ' e^{3\psi - 4\nu}$ is a dielectric displacement.  
The flux of electric field in this case is 
\begin{equation}
\mathbf{\Phi} = \mathcal D_r \times S = 
4\pi a \omega ' e^{3\psi - 4\nu} = 4 \pi q .
\label{sec1-17}
\end{equation}
Consequently we can identify the constant $q$ with an electric charge. 
One can note that it is not the point charge located in some 
point but this constant is an amount of the electric force lines. 
In Ref. \cite{dzhsin} was shown that the solutions of 
the 5D Kaluza-Klein equations with this metric have the following 
qualitative behaviour 
\begin{enumerate} 
\item 
$0 \leq H_{KK} < E_{KK}$ (or $q > Q$). The solution is 
\textit{the regular finite flux tube}. 
The throat between the surfaces at $\pm r_0$ is filled with both 
``electric'' and ``magnetic'' fields. The longitudinal
distance between the $\pm r_0$ surfaces depends on the relation 
between electric and magnetic fields.  
\item 
$H_{KK} = E_{KK}$ (or $q = Q$). In this case the solution is 
\textit{\textit{an infinite flux tube}} filled
with constant electric and magnetic fields. The cross-sectional
size of this solution is constant ($a=$ const.). 
\item 
$0 \leq E_{KK} < H_{KK}$ (or $q < Q$). In this case we have 
\textit{a singular finite 
flux tube} located between two (+) and (-) electrical and magnetic  
charges located at $\pm r_0$. Thus the longitudinal 
size of this object is again finite, but now the cross
sectional size decreases as $r \rightarrow r_0$. At
$r = \pm r_0$ this solution has real singularities which
we interpret as the locations of the charges. 
\end{enumerate} 
Here $r_0$ defines the places where $ds^2 = 0$ 
($\chi, \theta , \varphi = \mathrm{const}$, $r = \pm r_0$). 
Let us introduce a parameter $\delta = 1 - Q/q$. For us is interesting 
the first kind of solutions : finite and infinite flux tube for which 
we have 
$\delta \geq 0$. We can consider (for $\delta > 0$, $\delta \neq 0$) 
the limited region ($|r| \leq r_0$)of this solution for which 
$a \approx l^2_{Pl}$. In Ref. \cite{dzh2} it is shown that this part 
($|r| \leq r_0$) of the flux tube solution ($|r| \leq \infty$) 
can be considered as a thin flux tube filled with the elctric and 
magnetic fields. According to the above mentioned reasonings 
it is a string-like object, for the care we can call his as a thread. 
\par 
It is necessary to note that one can insert this part of flux tube 
solution between 
two Reissner-Nordstr\"om black holes \cite{dzh1}. This situation is much 
more interesting because it is like to string attached to 2 $D-$branes : 
the flux tube solution is the string, each Reissner-Nordstr\"om solution 
is $D-$brane and joining on the event horizon takes place. In this 
approach the string (dimensionally reduced flux tube solution) has a lightlike 
world sheet boundaries what is very close to $H-$branes approach 
introduced in Ref.\cite{kogan}. 

\section{Reduction of Flux Tube to String}

Now we would like to reduce our initial 5D Lagrangian to 2D Lagrangian. 
At this step we set the sizes of 5$^{\mathrm{th}}$ and $S^2$ dimensions 
$\approx l_{Pl}$.  The first step is the usual 5D $\rightarrow$ 4D 
Kaluza-Klein dimensional reduction. Following, for example, to review 
\cite{wesson} we have  
\begin{equation}
\frac{1}{16 \pi \stackrel{(5)}{G}}  \stackrel{(5)}{R} = 
\frac{1}{16 \pi G} \stackrel{(4)}{R} - \frac{1}{4} \phi^2 
F_{\mu \nu} F^{\mu \nu} + \frac{2}{3} 
\frac{\partial_\alpha \phi \partial^\alpha \phi}{\phi^2}
\label{sec2-10}
\end{equation}
where $\stackrel{(5)}{G} = G \int dx^5$ is 5D gravitational constant; 
$G$ is 4D gravitational constant; 
$\stackrel{(5),(4)}{R}$ are 5D and 4D Ricci scalars respectively. 
The 5D metric has the following form 
\begin{equation}
d \stackrel{(5)}{s^2} = g_{AB} dx^A dx^B = g_{\mu \nu}(x^\alpha)dx^\mu dx^\nu - 
\phi^2(x^\alpha)
\left (
dx^5 + A_\mu(x^\alpha) dx^\mu
\right )^2 
\label{sec2-20}
\end{equation}
where $A,B = 0,1,2,3,5$ are 5D indices; $\mu ,\nu = 0,1,2,3$ are 
4D indices. The determinant of 5D and 4D metrics are 
connected as 
$\stackrel{(5)}{g} = \stackrel{(4)}{g}\phi$. 
We assume that the length of 5$^{\mathrm{th}}$ dimension is $\approx l_{Pl}$. 
One of the paradigm of quantum gravity says us that it is a minimal 
length in the Nature. Physically it means that not any physical fields 
can have any structure inside this region. For us it means that these 
fields $f(x^\mu) \approx f(x^\mu + \delta x^\mu)$ if 
$\delta x^\mu \leq l_{Pl}$. These arguments allow us to say that the 
physical fields are not depend on the 5$^{\mathrm{th}}$ coordinate (of coarse, 
if its length is $\approx l_{Pl}$) and consequently 
$\partial f/\partial x^\mu \approx 0$ inside Planck scale. 
\par 
The next step is reduction from 4D to 2D. Let us remember that we consider 
the region of spacetime where the topology is 
$M^2 \times S^2 \times S^1$ and the linear sizes of $S^2$ are 
$\approx l_{Pl}$. As above we can conclude that all physical fields can not 
depend on the coordinates on the sphere $S^2$. For an external 
observer it is a physical (not mathematical !) point. Our statement 
is that \textit{this point can not be colored by different colors}. 
The 4D metric can be expressed as 
\begin{eqnarray}
d\stackrel{(4)}{s^2} & = & g_{\mu \nu} dx^\mu dx^\nu = 
g_{ab}(x^c) dx^a dx^b + 
\nonumber \\
&&\chi(x^c)
\left (
\omega ^{\bar i} + B^{\bar i}_a(x^c) dx^a
\right )
\left (
\omega _{\bar i} + B_{\bar i a}(x^c) dx^a
\right )
\label{sec2-30}
\end{eqnarray}
where $a,b = 0,1$; $x^a$ are the time and longitudinal coordinates; 
$-\omega^{\bar i} \omega_{\bar i} = dl^2$ is the metric 
on the 2D sphere $S^2$; all physical quantities 
$g_{ab}, \chi$ and $B_{\bar i a}$ can depend only on the 
physical coordinates $x^a$. Accoridingly to Ref. \cite{coq} we have 
the following dimensional reduction to 2 dimensions 
\begin{eqnarray}
\stackrel{(4)}{R} & = & \stackrel{(2)}{R} + R(S^2) - 
\frac{1}{4} \Phi^{\bar i}_{ab} \Phi^{ab}_{\bar i} - 
\nonumber \\
&&\frac{1}{2} h^{ij}h^{kl}
\left (
D_a h_{ik} D^a h_{jl} + D_a h_{ij} D^a h^{kl}
\right ) - 
\nabla^a 
\left (
h^{ij} D_a h_{ij}
\right )
\label{sec2-40}
\end{eqnarray}
where $\stackrel{(2)}{R}$ is the Ricci scalar of 2D spacetime; 
$D_\mu$ and $\Phi^{\bar i}_{ab}$ are, respecively, the covariant derivative 
and the curvature of the principal connection $B^{\bar i}_a$ and 
$R(S^2)$ is the Ricci scalar of the sphere 
$S^2 =  \mathrm{SU(2)/U(1)}$ with 
linear sizes $\approx l_{Pl}$; $h_{ij}$ is the metric on $\mathcal L$ 
\begin{eqnarray}
\mathrm{su}(2) & = & \mathrm{u}(1) \oplus \mathcal L ,
\label{sec2-50}\\
\mathrm{su(2)} & = & \mathrm{Lie (SU(2))} ,
\label{sec2-60}\\
\mathrm{u}(1) & = &\mathrm{Lie(U(1))}
\label{sec2-70}
\end{eqnarray}
here $\mathcal L$ is the orthohonal complement of u(1) algebra in su(2) 
algebra; the index $i \in \mathcal L$. The metric $h_{ij}$ is 
proportional to the scalar $\chi$ in Eq. \eqref{sec2-30}. 
\par 
Now we would like to consider the situation with the electromagnetic 
fields $A_\mu$ and $F_{\mu \nu}$. 
\begin{eqnarray}
A_\mu & = & \left \{ A_a, A_i \right \} \; 
A_a \; \text{is the vector;} \; 
A_i \; \text{are 2 scalars}; 
\label{sec2-80}\\
F_{ab} & = & \partial_a A_b - \partial_b A_a \; 
\text{is the Maxwell tensor for} \; A_a;
\label{sec2-90}\\
F_{ai} & = & \partial_a A_i ,
\label{sec2-100}\\
F_{ij} & = & 0
\end{eqnarray}
here we took into account that $\partial_i = 0$ as \textit{the point 
cannot be colored}. 
\par 
Connecting all results we see that only the following 
\textit{physical fields} on the our string-like object (thread) are possible : 
2D metric $g_{ab}$, gauge fields $B^{\bar i}_a$, vectors $A_a$, 
tensors $F_{ab}, F_{a\bar i}, \Phi^{\bar i}_{ab}$, scalars $\chi$ and 
$\phi$. 

\section{Discussion}

The gravitatonal flux tube solutions with the cross section 
on the Planck level 
are interesting objects. In the longitudinal direction they have classical 
properties but in the transversal directions they become quantum mechanical. 
In fact, it is some mixture of classical and quantum characteristics 
in one object~! 
For example, on the thread we can investigate such extra complicated quantum 
object as spacetime foam only in two dimensions. Also in the case of small 
perturbations on the background flux tube metric the transport of energy 
through this string-like object from one part of Universe to another 
becomes possible and so on. 
\par 
Another interesting property in this approach is that the 
string-like objects are the objects in the spirit of the Kaluza - Klein 
paradigm in gravity, \textit{i.e.} a dimensional reduction of some initial 
high dimensional pure gravitational spacetime to another low-dimensional 
one. 

\section{Acknowledgment}
VD is very grateful to the DAAD for the financial support and 
to the Alexander von Humboldt Foundation for the futher support. 
He thanks Prof. H. Kleinert for hospitality in his research group.


\begin{thebibliography}{99}

\bibitem{dzh2}
V. Dzhunushaliev, ``Strings in the Einstein paradigm of matter'', 
gr-qc/0205055. 

\bibitem{pol}
A. Polyakov,  Nucl. Phys., \textbf{B268}, 406 (1986).

\bibitem{kleinert}
H. Kleinert, Phys. Lett., \textbf{B174}, 335 (1986).

\bibitem{dzhsin}
V. Dzhunushaliev and D. Singleton, Phys. Rev. \textbf{D59},
064018 (1999).

\bibitem{dzh1}
V. Dzhunushaliev, Mod. Phys. Lett., \textbf{A13}, 2179 (1998).

\bibitem{kogan} 
Ian I. Kogan and Nuno B.B. Reis, Int. J. Mod. Phys., 
\textbf{16}, 4567 (2001).

\bibitem{wesson}
J. M. Overduin and P. S. Wesson, Phys. Rept., \textbf{283}, 303 (1997).

\bibitem{coq}
R. Coquereaux and A. Jadczyk, Commun. Math. Phys., \textbf{90}, 79 (1983). 

\end{thebibliography}
\end{document}